%% file: text_final.tex
\documentstyle[11pt,newpasp,twoside,epsf]{article}
\def\edcomment#1{\iffalse\marginpar{\raggedright\sl#1\/}\else\relax\fi}
\reversemarginpar

\include{texdefs}

\begin{document}
\title{{\boldmath $H\!ST$} STIS Ultraviolet 
Spectral Evidence for Outflows in Extreme
Narrow-Line Seyfert 1 Galaxies}
 \author{Karen M. Leighly}
\affil{Columbia Astrophysics Laboratory, Columbia University, 550
 W.\ 120th Street, New York, NY 10027; and
Department of Physics and Astronomy, 
The University of Oklahoma, 440 W.\ Brooks
St., Norman OK 73019}

\begin{abstract}
I present and discuss the results of \HST\ STIS observations of 
IRAS 13224$-$3809 and 1H~0707$-$495, two
narrow-line Seyfert 1 (NLS1) galaxies.
We discovered that high-ionization UV emission lines are much broader
and strongly blueshifted compared with the low-ionization and 
intermediate-ion\-iza\-tion lines, which are relatively narrow and 
centered at the rest
wavelength.  We interpret this as evidence that the high-ionization
lines come from a wind, while the low-ionization lines are emitted
from the accretion disk or low-velocity base of the wind.  The
optically thick disk blocks our view of the receding wind.

We also found that not all NLS1s display strongly blueshifted emission
lines, and the degree of asymmetry is inversely correlated with the
equivalent width and correlated with the Si\,{\sc iii}] to
\ciii] ratio, a density indicator.  It may also be significant
that these NLS1s display the extreme in the X-ray properties: they
have the strongest X-ray soft-excess components and the 
highest-amplitude, flaring variability.  We postulate that this combination of
properties are all related to a high value of $L/M_{\rm BH}$, even among NLS1s.

The high-ionization and low-ionization 
lines are nearly completely kinematically
separated, a fact that allows us to study the conditions in the disk
and wind separately.  The ratios and equivalent widths of the emission
lines produced in the wind, especially the comparative amounts of
\ciii], \civ, and the 1400\,\AA\ feature, are somewhat
challenging to explain.  To illustrate this, we consider the simple,
but probably unrealistic, single-zone model in which the line-emitting
gas sees the same continuum that we see.  We find two solutions: one
in which the gas is so dense that \civ\ is thermalized, and
another in which the \civ\ is reduced because of a relatively
higher optical depth in gas with a velocity gradient.

\end{abstract}

\section{Introduction}

Broad optical and UV emission lines are a nearly ubiquitous and
identifying feature of active galactic nuclei (AGN) spectra.  Yet
despite more than thirty years of intensive study, many very basic
properties are not well constrained or understood, including the
geometry of the emission-line region, the origin of the inferred high
velocities, and the physical conditions in the line-emitting gas.

Within the last five years it has become well-established that a
subclass of AGN known as narrow-line Seyfert 1 (NLS1) galaxies exhibit a
characteristic set of X-ray properties including high-amplitude and
rapid X-ray variability, ultrasteep X-ray spectra, and often a strong
soft-excess component (e.g., Leighly 1999a, b).  These properties are
interpreted as evidence that NLS1s are accreting at a higher fraction
of the Eddington rate.  Because all AGN are believed to be powered by
a supermassive black hole, this result is fundamental.

This discovery represents the first identification of a direct link
between the central engine and the kinematics of the
emission-line region.   Much of the new work done has
focused on the influence of the inferred different SED that NLS1s have
on the distance to the emission-line region (Wandel \& Boller 1998)
and the density in terms of an ionization instability (Kuraszkiewicz et
al.\ 2000), or on the usual method of reverberation mapping.  The
detailed study of NLS1 emission line profiles has not yet been been
performed.  We show below that because of the peculiarities of some NLS1
profiles, this method may prove significantly more powerful than
usually thought.

\section{{\boldmath $H\!ST$ UV STIS Spectra of
IRAS 13224$-$3809 and 1H 0707$-$495}}

We observed the NLS1 galaxies IRAS 13224$-$3809 and
1H 0707$-$495 using the \HST\ STIS in June and February 1999,
respectively.  The spectra are shown in Fig.\ 1.  The spectra are as
blue as that of the average quasar (e.g., Leighly 2000; Leighly \&
Halpern 2000), and all of the absorption lines present in the
spectra originate in our Galaxy.  The spectra are practically
identical to each other and strongly resemble that of the NLS1 I~Zw~1
(Laor et al.\ 1997), indicating that these spectra represent not an
isolated phenomenon but rather reflect characteristic behavior
resulting from some set of physical conditions.

\begin{figure}
\plotone{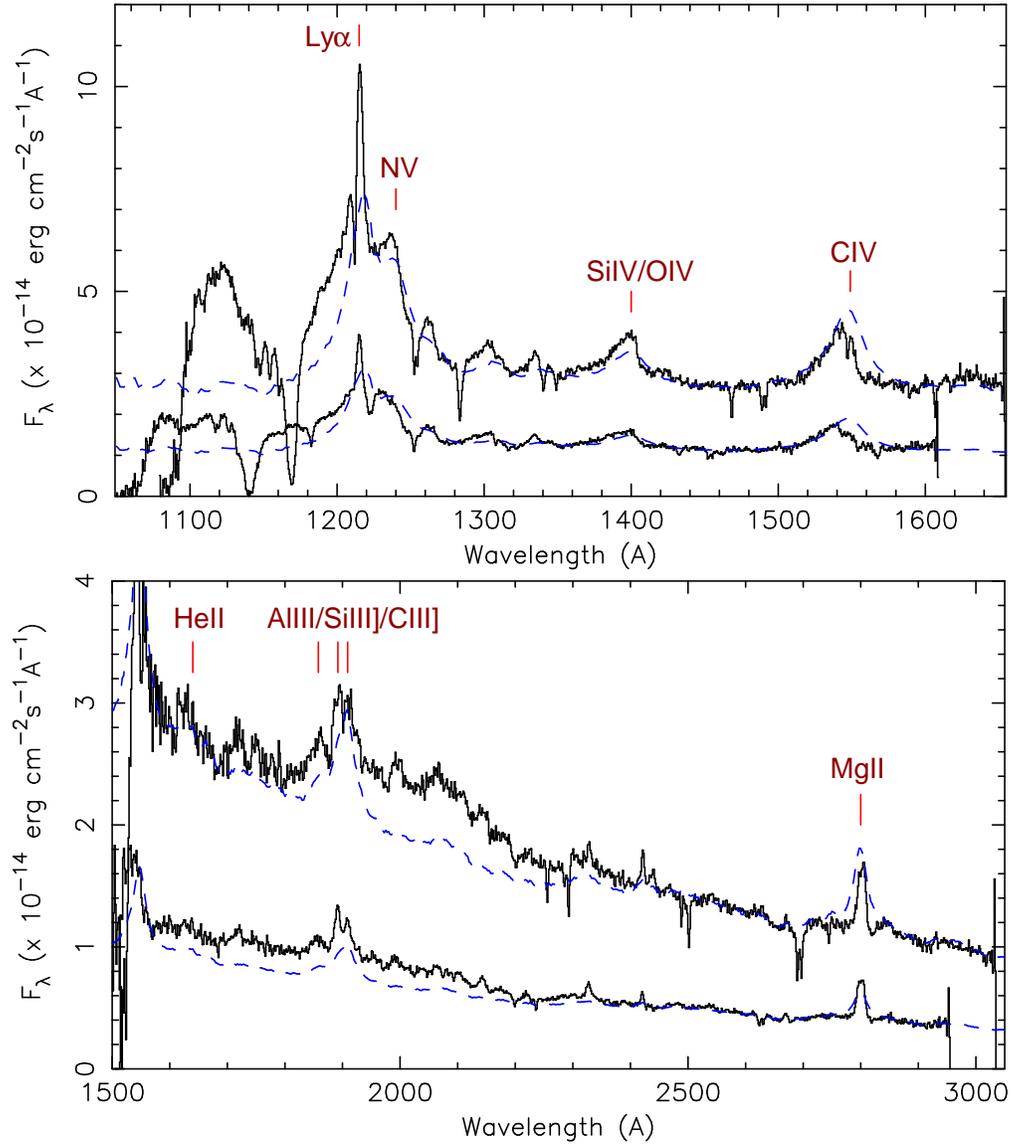}
\caption[]{The solid lines show \HST\ STIS UV spectra of the NLS1s
1H~0707$-$495
(upper) and IRAS~13224$-$3809 (lower).  The dashed lines show the
rescaled average quasar spectrum (Francis et al.\ 1991).  The rest
wavelengths of the principal lines used for analysis are marked. All
of the absorption lines originate in our Galaxy.}
\end{figure}

The most remarkable features of these spectra are the emission-line
profiles.  We find that the high-ionization lines, including
\civ\,$\lambda 1549$, Ly$\alpha\,\lambda 1215$,
\nv\,$\lambda 1240$, and the 1400\,\AA\ feature comprised of
Si\,{\sc iv} and O\,{\sc iv} are much broader than the low-ionization and
intermediate-ionization lines, including \mgii\,$\lambda 2800$,
Al\,{\sc iii}\,$\lambda\lambda 1855$, 1863, Si\,{\sc iii}]\,$\lambda
1892$ and \ciii]\,$\lambda 1909$ (Fig.\ 2).
The trend for the high-ionization lines to be broader 
than the low-ionization lines has been observed before, 
but these spectra represent
an extreme of the phenomenon.  The high-ionization lines are also
strongly blueshifted with respect to the low-ionization lines. This
trend has also been previously observed (e.g., Tytler \& Fan 1992;
Marziani et al.\ 1996), but these lines again represent the extreme.

\begin{figure}
\plotone{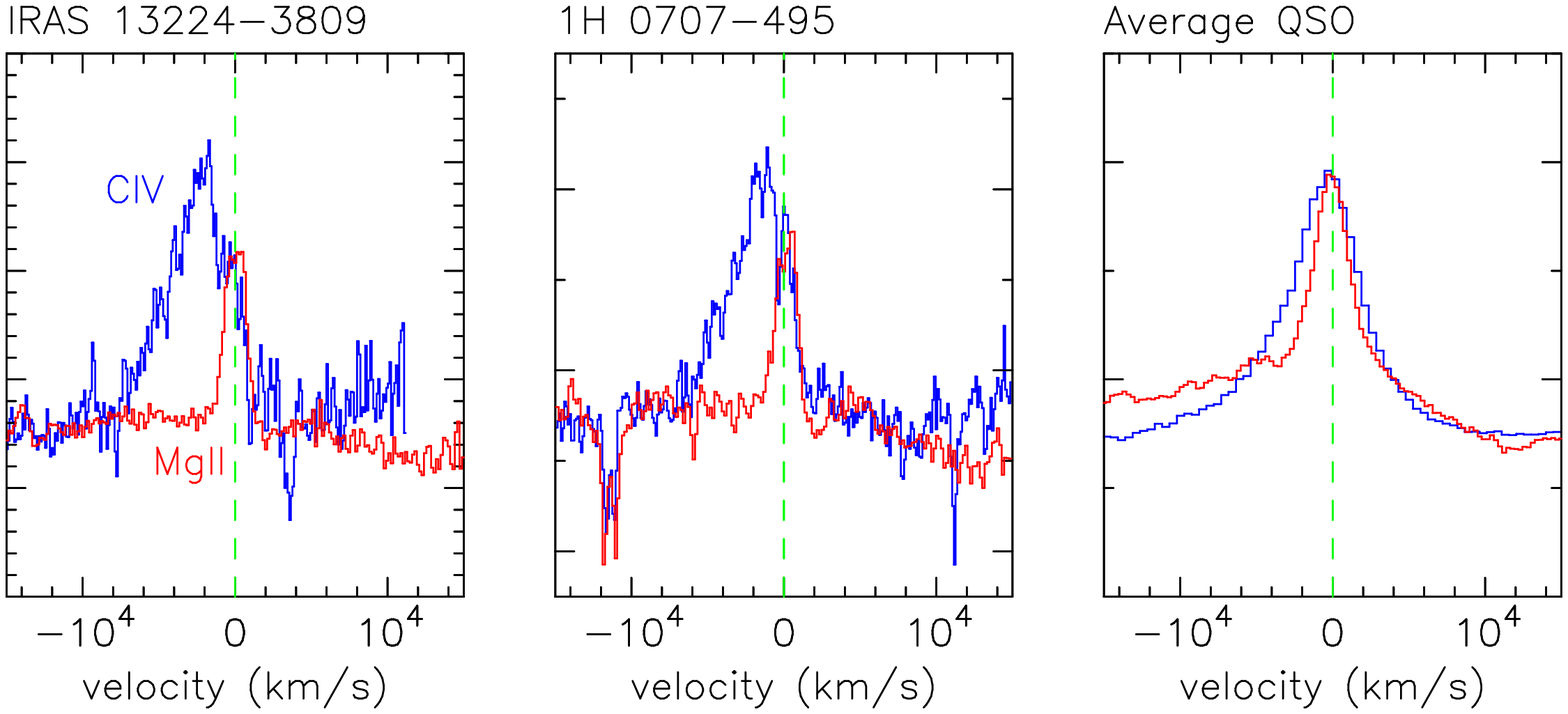}
\caption[]{The rescaled representative high-ionization line
\civ\ superimposed on the representative low-ionization line
\mgii\  as a function of velocity, for our two NLS1s and the
average quasar (Francis et al.\ 1991).  The average quasar
high-ionization line is slightly broader and slightly blueshifted
compared with the low-ionization line.  In contrast, the
low-ionization lines in NLS1s are much narrower and the
high-ionization lines are strongly blueshifted.}
\end{figure}

The most straightforward interpretation of this result is that the
high-ionization lines are produced in a wind, perhaps accelerated by
radiative line driving (e.g., Proga, Stone, \& Kallman 2000), while
the low-ionization lines are produced in the accretion disk or
low-velocity base of the wind.  The disk is optically thick, so the
emission from the receding wind is not seen. Disk--wind models for AGN
emission lines have been discussed previously (Collin-Souffrin et al.\
1988); however, these spectra arguably provide one of the strongest
pieces of evidence supporting this model and ruling out other models.

The X-ray properties of narrow-line Seyfert 1 galaxies, including
their enhanced X-ray variability, have led to the interpretation that
NLS1s have a smaller black-hole mass relative to their accretion rate
(e.g., Leighly 1999a, b, and references therein).  The fact that we find
evidence for emission from high-velocity winds in NLS1s rather than in
other AGN is intuitively sensible, since a relatively high-luminosity
to black-hole mass ratio may very well produce a powerful wind.  Many
NLS1s have also been found to have very high Si\,{\sc iii}] to
\ciii] ratios, an indication of high densities (e.g., Wills et
al.\ 1999).  High densities in the accretion disk may also be expected
when the accretion rate is high.

Interestingly, however, not all NLS1s show these highly blueshifted
high-ionization emission lines or high Si\,{\sc iii}] to \ciii]
ratios.  The high-ionization lines of some NLS1s, while broader than
the respective low-ionization lines, are still relatively narrow and
symmetric.  We discovered that the \civ\ equivalent width is
inversely correlated with the line asymmetry, as shown in Fig.\
3\footnote{We note that this result cannot be attributed to a Baldwin
effect, since our objects have a median luminosity compared with the
sample of NLS1s with \HST\ spectra.}.  We postulate that this result can
be qualitatively explained if the high-ionization lines are considered
to be approximately comprised of two components: one from the outflow
and the other from the lower-velocity material at the base of the wind or
from the accretion disk. The highly asymmetrical, low equivalent width
emission lines are observed when only the wind component is seen and
the emission lines are dominated by the wind.  This may be because the
wind is so thick that it blocks the more energetic photoionizing
photons from reaching the low-velocity material at the base of the
wind.  The Si\,{\sc iii}] to \ciii] ratio is similarly
correlated with the line asymmetry.

\begin{figure}
\plotone{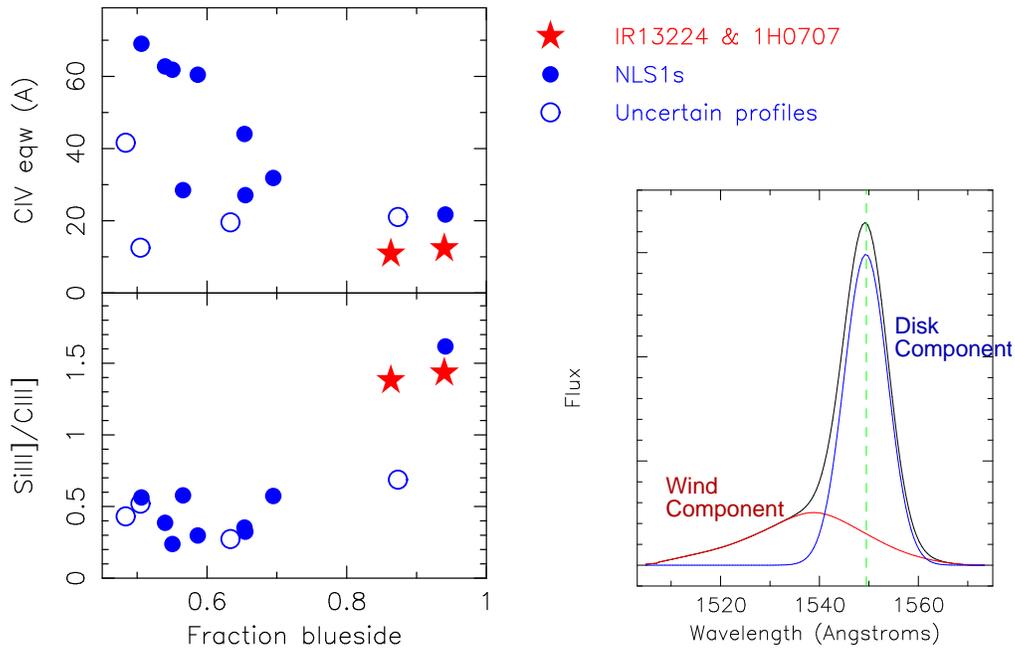}
\caption[]{{\it Left:} The \civ\ line asymmetry, parameterized
simply as the fraction of the emission line blueward of the rest
wavelength, plotted against the equivalent width and the
Si\,{\sc iii}] to \ciii] ratio, a density indicator.  The stars mark
IRAS~13224$-$3809 and 1H~0707$-$495, the subjects of detailed
analysis (Leighly \& Halpern 2000).  Some of the objects had
associated absorption and the reconstructed profiles are uncertain
(circles). {\it Right:} A schematic diagram of the decomposition of a
high-ionization line in most NLS1s; our objects show only the wind component.}
\end{figure}

Why do IRAS~13224$-$3809 and 1H~0707$-$495 show these extreme line
profiles and other NLS1s do not?  It is probably important that these
two objects are not run-of-the-mill NLS1s with respect to their X-ray
properties.  IRAS~13224$-$3809 and 1H~0707$-$495 fall at the extreme
of the correlation between high amplitude X-ray variability 
and soft-excess strength observed in the \ASCA\ sample of NLS1s (Leighly
1999a, b). They also have unusual features around 1\,keV that have been
interpreted as evidence for absorption in a high-velocity outflow
(Leighly et al.\ 1997), although we note that no evidence for similar
absorption was detected in the UV.

Based on these results, we are developing the following paradigm
(Leighly \& Halpern 2000): in concurrence with other investigators, we
support the idea that all NLS1s have a higher $L/M_{\rm BH}$ ratio than do
Seyfert 1 galaxies with broad optical lines.  However, some NLS1s have
a higher $L/M_{\rm BH}$ ratio than others.  The objects that have the
highest $L/M_{\rm BH}$ ratios are accreting close to the Eddington rate
and suffer accretion instabilities, a condition manifesting itself in
high-amplitude, flaring X-ray variability.  It is in these objects
that are close to the Eddington limit that we also observe 
high-ionization lines that are dominated by emission from the wind.

\begin{figure}
\plotfiddle{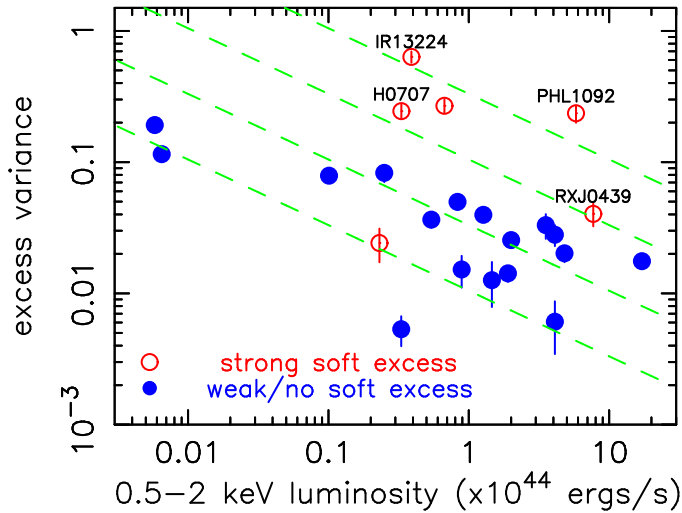}{180pt}{0}{70}{70}{-131}{0}
\caption[]{The objects that show wind-dominated high-ionization UV
emission lines (IRAS~13224$-$3809 and 1H~0707$-$495) also have the
highest excess variance at a particular X-ray luminosity and the
overall steepest X-ray spectra (strong soft excess).  These objects
are outliers from the expected variance--luminosity correlation
denoted by dashed lines and we think that these objects are typified
by an exceptionally high $L/M_{\rm BH}$ ratio.  Also marked are two other
NLS1s that show high-amplitude, flaring variability and strong soft
excesses.  If our paradigm is correct, these two objects should also
show wind-dominated high-ionization lines.}
\end{figure}

\section{Photoionization Modeling}

Analysis of the ultraviolet emission lines in IRAS~13224$-$3809 and
1H~0707$-$495 promises great rewards. The high-ionization and low-ionization
lines are nearly completely kinematically separated, so the conditions
in the wind and disk\footnote{We do not know with certainty the
geometrical and physical origin of the emission lines in the objects
we are discussing here.  However, for simplicity, we refer to the
highly blueshifted high-ionization lines as originating in the
``wind'', and the narrow, symmetric low-ionization lines as
originating in the ``disk''. These distinctions are somewhat similar
to the HIL and LIL regions previously proposed (Collin-Souffrin et
al.\ 1988), except that \ciii] and other similar
intermediate-ionization lines appear to be produced in the disk.} can
each be determined in principle.  This kind of component analysis has
also been discussed in Baldwin et al.\ 1996; our spectra may provide
arguably better constraints because the kinematic separation is
cleaner.

We first try to model the blueshifted high-ionization emission lines
from the wind using one-zone models, although we acknowledge that such
models almost certainly must be inadequate. We first assume that the
emitting gas sees the same continuum that we do, although this may
also be unrealistic, as discussed below. We find that the following
observed properties place the most severe constraints on the models:

\begin{enumerate}
\item The \civ\ equivalent width is low relative
to other emission lines in these objects compared with other AGN.
\item The  1400\,\AA\/ feature is stronger in these objects compared
with other AGN.  This feature is most probably comprised of 
Si\,{\sc iv} and O\,{\sc iv}, and profile deconvolution indicates that
much of it must come from the wind.  
\item There is no evidence for a broad, blueshifted component of
\ciii] or any other intermediate-ionization or low-ionization line.  We
nominally set the upper limit of the equivalent widths contribution
to the wind from these lines to be 1\,\AA.
\end{enumerate}

We find two scenarios that roughly fulfill these constraints; the
details are discussed by Leighly \& Halpern (2000).  The first scenario
assumes that the gas is very dense.  In this case, the \ciii]
equivalent width is reduced as the C$^{+2}$ ions become collisionally
de-excited before emission occurs.  If the gas is denser still,
\civ\ becomes thermalized;  \civ\ is the dominant coolant
over a wide range of ionization parameters, so when it becomes
thermalized, the cooling must shift to other emission lines such as
Si\,{\sc iv}.  This is somewhat similar to the scenario proposed by
Kuraszkiewicz et al.\ (2000) from an analysis of NLS1 line ratios
without the benefit of kinematic decomposition.  We found that this
process becomes effective when the density is $10^{13.5}\,\rm cm^{-3}$.
We find that the equivalent widths and line ratios are explained when
the ionization parameter
$\log U =-1.5$, the column density is $10^{21}\,\rm cm^{-2}$, and
the covering fraction is 0.25.  It is important to note that in this
case the ``wind'' must have approximately three orders of magnitude
higher density than the ``disk,'' since we observe \ciii] to be
narrow and symmetric and therefore present among the disk lines.  We
note that nitrogen must be overabundant, as is often inferred in AGN
(e.g., Hamann \& Ferland 1999).

\begin{figure}
\plotone{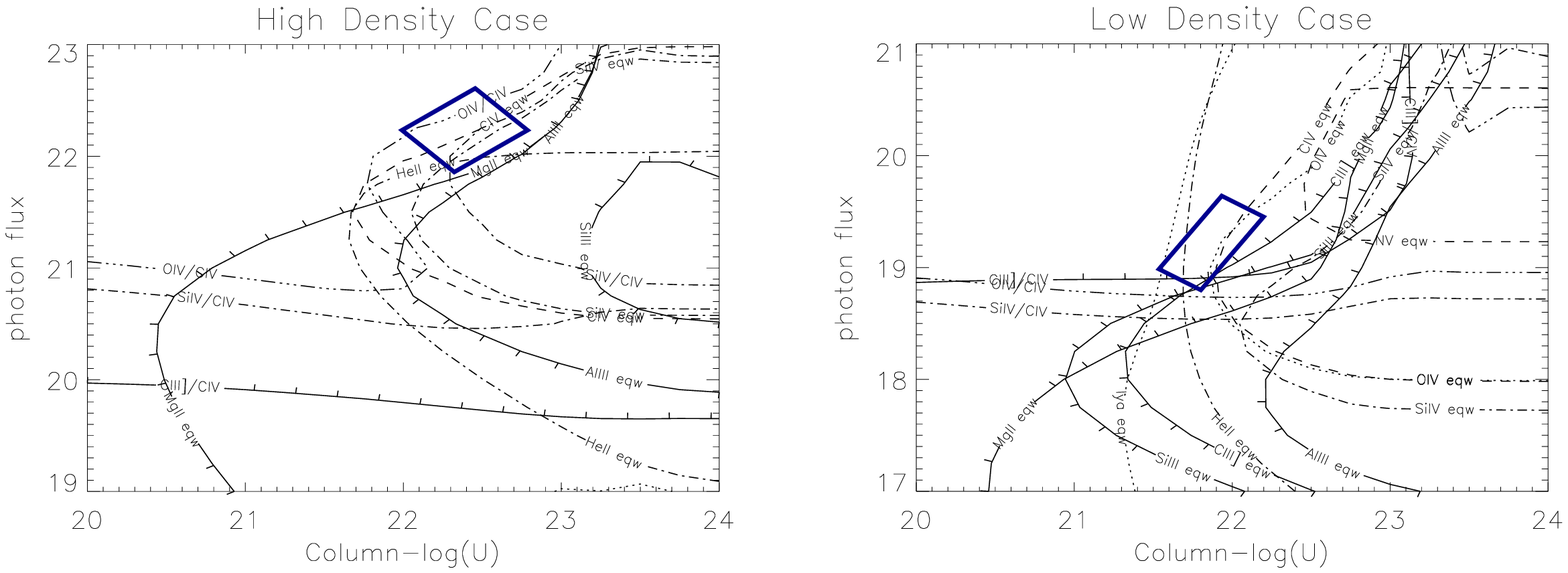}
\caption[]{Results of one-zone photoionization modeling of the wind
emission lines, assuming that the line-emitting gas sees the same
continuum that we see.  Contours mark observed equivalent widths and
line ratios, while solid lines with tick marks indicate equivalent
width upper limits.  {\it Left:} The high-density solution
($n=10^{13.5}\rm cm^{-3}$), in which the \civ\ is thermalized,
reducing the equivalent width relative to Si\,{\sc iv}. {\it Right:}
The low-density solution ($n=10^{10.5}\,\rm cm^{-3}$), in which
\civ\ equivalent width has been reduced by a factor of 4 to
approximate the effect of radiative transfer in moving material.}
\end{figure}
 
We found a second solution to the problem.  If the wind is smooth and
if there are significant velocity gradients in the wind, the radiative
transfer of the lines will be affected.  This is a well-known
phenomenon in stellar atmospheres that potentially affects emission
lines in other windy systems such as supernovae, CVs, and hot stars.
Under the Sobolev approximation, the optical depth to the line depends
on a number of factors including the velocity gradient that will
affect all resonance lines the same way. However, the optical depth
also has some dependence on the atomic physics that will result in a
different opacity for different lines.  The primary factor is the
product of the line absorption oscillator strength, the
atomic abundance, and the wavelength of the emission line.  This
factor is about 4--5 times larger for \civ\ than it is for the
other high-ionization resonance lines, and much larger than it is for
semiforbidden lines such as O\,{\sc iv}.  When the scattering optical
depth is large, the escape probability will be proportional to
$1/\tau$ in the Sobolev approximation (Rybicki \& Hummer 1983).
Taking this difference in opacity into account, we find that we can
find an approximate solution for ionization parameter $\log(U)=-1.5$,
density $n=10^{10.5}\,\rm cm^{-3}$, column density $10^{21}\,\rm cm^{-2}$,
and a covering fraction of 0.5.  Again, nitrogen should be overabundant.

It is interesting to note that the effects of differential opacity due
to velocity gradients may be present also in the disk emission lines.
We observe fairly strong features at least partially attributable to
Si\,{\sc ii}\,$\lambda 1260$, Si\,{\sc ii}\,$\lambda 1305$, and
\cii\,$\lambda 1335$ (Fig.\ 1).  Such strong Si\,{\sc ii}
features have also been noted in the NLS1 archetype I~Zw~1 (Laor et
al.\ 1997) and other narrow-line quasars (Baldwin et al.\ 1996).
Recently, Bottorff et al.\ (2000)
have shown that these emission lines are
predicted to be enhanced under conditions of turbulence or velocity
gradients.

However, there are a  number of factors that could strongly affect the
predicted emission lines that have not been accounted for  here:
\begin{enumerate}
\item  The one-zone model is certainly overly simplistic.  Spatial
ionization stratification may be present among the high-ionization
lines although we note that we obtain a fairly good fit to other
high-ionization lines using a template constructed from the
\civ\ profile.  
\item Is the wind smooth or clumpy?  Radiative transfer of \civ\
relative to other lines 
may or may not be a significant factor in a two-phase wind, depending
on the density and ionization contrasts.
\item These objects are among the most X-ray variable NLS1s, and
therefore the effects of a rapidly-variable ionizing continuum could
be important.  It is easy to show that the average line emission
produced from two continua is not the same as the line emission from
the average continuum.  The wind is likely to be supersonic, and if
the wind is accelerated by line-driving, variability may produce
shocks which could change the density and the ionization.
\item What are the effects of radiative transfer?  A
zeroth-order discussion is presented here, and any attempt 
to couple radiative transfer to the line emission 
obtained from CLOUDY (Ferland 1997) is necessarily
approximate because the radiative transfer will affect the cooling as
well as the level populations.
\item Is the continuum incident upon the line emitting material the
same continuum that we see?  Models for AGN winds employing radiation
line-driving require a screen of highly ionized gas between the
central engine and the line-emitting wind to absorb the soft X-rays
and prevent the wind gas from becoming over-ionized (Murray et al.\
1995; Proga, Stone \& Kallman 2000). There are also some indications
that the soft X-rays in these highly variable objects may be beamed
(Brandt et al.\ 1999).
\end{enumerate}

\end{document}

%% file: texdefs.tex
\def\HST{{\it HST}}

\def\ASCA{{\it ASCA}}

\def\arcsec{\ifmmode '' \else $''$\fi}
\def\arcmin{\ifmmode ' \else $'$\fi}
\def\arcsecpoint{\ifmmode ''\!. \else $''\!.$\fi}
\def\arcminpoint{\ifmmode '\!. \else $'\!.$\fi}
\def\kms{\ifmmode {\rm km\ s}^{-1} \else km s$^{-1}$\fi}
\def\Hubble{\ifmmode {\rm km\ s}^{-1}\ {\rm Mpc}^{-1} 
	\else km s$^{-1}$ Mpc$^{-1}$\fi}
\def\ergsec{\ifmmode {\rm ergs\ s}^{-1} \else ergs s$^{-1}$\fi}
\def\eflux{\ifmmode {\rm ergs\ s}^{-1}\;{\rm cm}^{-2}
	  \else ergs s$^{-1}$ cm$^{-2}$\fi}
\def\efluxA{\ifmmode {\rm ergs\ s}^{-1}\;{\rm cm}^{-2}\;{\rm \AA}^{-1}
	  \else ergs s$^{-1}$ cm$^{-2}$ \AA$^{-1}$\fi}
\def\efluxHz{\ifmmode {\rm ergs\ s}^{-1}\;{\rm cm}^{-2}\;{\rm Hz}^{-1}
	  \else ergs s$^{-1}$ cm$^{-2}$ Hz$^{-1}$\fi}
\def\cc{\ifmmode {\rm cm}^{-3} \else cm$^{-3}$\fi}
\def\vFWHM{\ifmmode v_{\mbox{\tiny FWHM}} \else
            $v_{\mbox{\tiny FWHM}}$\fi}
\def\Msun{\ifmmode M_{\odot} \else $M_{\odot}$\fi}
\def\Lsun{\ifmmode L_{\odot} \else $L_{\odot}$\fi}
\def\Halpha{\ifmmode {\rm H}\alpha \else H$\alpha$\fi}
\def\Hbeta{\ifmmode {\rm H}\beta \else H$\beta$\fi}
\def\Hgamma{\ifmmode {\rm H}\gamma \else H$\gamma$\fi}
\def\Hdelta{\ifmmode {\rm H}\delta \else H$\delta$\fi}
\def\Lya{\ifmmode {\rm Ly}\alpha \else Ly$\alpha$\fi}
\def\Lyb{\ifmmode {\rm Ly}\beta \else Ly$\beta$\fi}
\def\hi{\ifmmode {\rm H}\,{\sc i} \else H\,{\sc i}\fi}
\def\hii{\ifmmode {\rm H}\,{\sc ii} \else H\,{\sc ii}\fi}

\def\cii{C\,{\sc ii}}
\def\ciii{\ifmmode {\rm C}\,{\sc iii} \else C\,{\sc iii}\fi}
\def\civ{\ifmmode {\rm C}\,{\sc iv} \else C\,{\sc iv}\fi}

\def\nv{N\,{\sc v}}

\def\mgii{Mg\,{\sc ii}}